\documentclass[a4paper,showpacs]{revtex4}  

\usepackage{amsmath,amssymb}  
\usepackage{graphicx} 
 
\newcommand{\ncd}{\newcommand}
\ncd{\e}{{\rm e}} \ncd{\ex}{{\rm e}}
\ncd\sgn{\mathop{\rm sgn}\nolimits}
\ncd\real{\mathop{\rm Re}\nolimits}
\ncd\imag{\mathop{\rm Im}\nolimits}
\ncd\Ker{\mathop{\rm Ker}\nolimits}
\ncd\sinc{\mathop{\rm sinc}\nolimits}
\ncd{\re}{\Re{e}} \ncd{\im}{\Im{m}}
\ncd\divergence{\mathop{\rm div}\nolimits}
\ncd\Tr{\mathop{\rm Tr}\nolimits}
\ncd\ts\textstyle
\ncd\arccosh{\mathop{\rm arccosh}\nolimits}
\ncd\arccot{\mathop{\rm arccot}\nolimits}
\ncd\arccoth{\mathop{\rm arccoth}\nolimits}
\ncd\Ei{\mathop{\rm Ei}\nolimits}  
\ncd{\Dtwo}{\mathcal{D}} 
\ncd{\Ttwo}{\mathcal{T}} 
\ncd{\lagomdot}{{\mbox{\large$\cdot$}}} 
\ncd{\dotprime}{^{(\lagomdot}{}'{}^)} 
\ncd{\nms}{\negmedspace}  
\ncd{\nts}{\negthickspace}  
\ncd{\mcl}[1]{\mathcal{#1}}  
\ncd{\beq} {\begin{equation}}  
\ncd{\eeq} {\end{equation}}  
\ncd{\BE} {\begin{eqnarray}}  
\ncd{\EE} {\end{eqnarray}}  
\ncd{\rarr} {\rightarrow}  
\ncd{\larr} {\leftarrow}  
\ncd{\lrarr} {\leftrightarrow}  
\ncd{\lbeq}[1]  {\label{eq: #1}}  
\ncd{\refeq}[1] {(\ref{eq: #1})}  
\ncd{\mrm}    {\mathrm}  
\ncd{\nn}{\nonumber}  
\ncd{\mbf}[1] {{\mathbf #1}}  
\ncd\T{\frac{1}{2}h^{\mu\nu}p_\mu p_\nu}  
\ncd{\ms}{\mathstyle}  
\ncd{\ds}{\displaystyle}  
\ncd{\bmth}[1] {\mbox{\boldmath $#1$}}  
\ncd{\abs}[1] {|#1|}  
\ncd{\ubold}{\mathbf u} 
\ncd{\Abold}{\mathbf A} 
\ncd{\Bbold}{\mathbf B} 
\ncd{\Mbold}{\mathbf M} 
\ncd{\tsfrac}[2]{{\ts\frac{#1}{#2}}}
\ncd{\dsfrac}[2]{{\displaystyle\frac{#1}{#2}}}
\ncd{\parfrac}[2]{\left(\frac{#1}{#2}\right)}
\ncd{\lagom}{\hspace{.6pt}} 
\ncd{\muk}{k} 
\ncd{\dumkonstant}{v_0} 
\ncd{\tdelta}{{\tilde\delta}} 
\ncd{\tnabla}{{\tilde\nabla}}
\ncd{\tbox}{{\tilde\square}} 
\ncd{\ok}{\quad\mathrm{\bf(ok!)}}
\ncd{\omegaAL}{\bar{\omega}}
\ncd{\omegaorig}{{}^{0\!}\omega}
\ncd{\omegaALorig}{{}^{0\!}\omegaAL}

\def\apriori{{\it a priori \/}} 
 \def\etal{{\it et al.\ }} \def\ie{{\it i.e.\ }}

\begin{document}

 
\title{Charged black holes in compactified spacetimes} 
  \author{Max Karlovini}
  \email{max@physics.muni.cz}
  \author{Rikard von Unge}
  \email{unge@physics.muni.cz}
  \affiliation{Department of Theoretical Physics and Astrophysics, Faculty of Science, Masaryk University, Kotla\u{r}sk\'a 2, 611 37 Brno, Czech Republic}

\vspace{2cm} 
 
\begin{abstract}{\normalsize
  We construct and investigate a compactified version of the
  four-dimensional Reissner-Nordstr\"om-NUT solution, generalizing the
  compactified Schwarzschild black hole that has been previously
  studied by several workers. Our approach to compactification is
  based on dimensional reduction with respect to the stationary
  Killing vector, resulting in three-dimensional gravity coupled to a
  nonlinear sigma model. Using that the original non-compactified
  solution corresponds to a target space geodesic, the problem can be
  linearized much in the same way as in the case of no electric nor
  NUT charge. An interesting feature of the solution family is that
  for nonzero electric charge but vanishing NUT charge, the solution
  has a curvature singularity on a torus that surrounds the event
  horizon, but this singularity is removed when the NUT charge is
  switched on. We also treat the Schwarzschild case in a more complete
  way than has been done previously. In particular, the asymptotic
  solution (the Levi-Civita solution with the height coordinate made
  periodic) has to our knowledge only been calculated up to a
  determination of the mass parameter. The periodic Levi-Civita
  solution contains three essential parameters, however, and the
  remaining two are explicitly calculated here.  }
\end{abstract} 
\vspace{.5cm} 

\maketitle
\section{Introduction}
String theory predicts that our world has more dimensions than the 4
we see around us every day. The usual way out of this apparent discrepancy
is to say that the additional dimensions are compactified and small. In the
simplest case the extra dimensions are just circles. String theory is also
a theory of gravity. It is therefore a natural and interesting question to
ask what is the behavior of the classical solutions of General Relativity in
spacetimes where one or more dimensions are compactified. This question is
however surprisingly hard to answer. The periodic analogue of a
Schwarzschild black hole in a space of topology $R3\times S1$ has been st
udied more or less independently by Majumdar \cite{Majumdar}, Papapetrou
\cite{Papapetrou}, Myers \cite{Myers}, Korotkin and Nicolai
\cite{Korotkin1,Korotkin2} as well as by Frolov and Frolov \cite{Frolov}. In
higher dimensions the full solution is known only implicitly 
\cite{HO1,HO3,HO4,Kol5,Kol6}.

The problem becomes even more interesting when one realizes, following
Gregory and Laflamme \cite{GL1,GL2}, that there is a competition between
different gravitational configurations with the same mass and charges but
with different symmetries and even of different horizon topology. Which
configuration is stable depends on the particular value of the mass and the
charges. For instance, a long and thin black string becomes unstable and
will "decay" into a new configuration with higher entropy. This new
configuration could for instance be a nontranslationally invariant black
string or an infinite array of black holes. The phase diagram of black
objects in higher dimensional circle space times has a very rich structure
\cite{Kol1,HO2,Kol3,HO5}. Only parts of it is accessible to analytical
methods and the most
interesting pices has been calculated only numerically
\cite{Wiseman1,Wiseman2,Kol2,Kol4,Wiseman3,Wiseman4} or using perturbation
theory \cite{Gubser}. It would therefore clearly be of great interest to
have a better analytical understanding of the various solutions that could
appear.

In this paper we perform a more modest task. We generalize the solution
given in \cite{Myers,Korotkin1,Frolov} to arbitrary electric and NUT
charge. In doing so we find a new parameter of the solution which has not
been recognized before. We analyze the physical meaning of this parameter.
We also notice that adding NUT charge to the solution makes it more well
behaved. Maybe this is a feature which would persist in higher dimensions?
Our method heavily relies on Weyl coordinates. However, as shown in
\cite{Emparan},
because of symmetry reasons, Weyl coordinates are expected to be useful at
most in 4 and 5 dimensions. This is something which would have to be
overcome if one would like to generalize our results to higher dimensions. 

This paper is organized as follows. Section 2 is a general discussion of
properties of the four dimensional Einstein-Maxwells equations important
for our problem. In Section 3 we discuss the (nonperiodic)
Reissner-Nordstr\"om-NUT solution. In Section 4 we find its periodic
generalization and discuss the metric in various limits. In Section 5 we
give our conclusions.

\section{Einstein-Maxwells equations for stationary spacetimes}
We write a general stationary metric with timelike Killing vector
$\xi^a$ as
\begin{equation}
  g_{ab} = -f^2\mu_a \mu_b + f^{-2}h_{ab},
\end{equation}
where
\begin{align}
  &f = \sqrt{-\xi^a\xi_a} \\
  &\mu_a = -f^{-2}\xi_a,
\end{align}
which implies that $h_{ab}\xi^b = 0$ and that $f$, $\mu_a$ and
$h_{ab}$ are stationary fields, \ie Lie dragged by $\xi^a$. We now
decompose the electromagnetic field strength $F_{ab}$ by defining the
electric and magnetic fields $E_a$ and $B_a$ with respect to $\xi^a$:
\begin{align}
  &E_a = F_{ab}\xi^b \\
  &B_a = *F_{ab}\xi^b, 
\end{align}
where $*F_{ab}$ is the dual field strength
\begin{equation}
  *F_{ab} = \tsfrac12\epsilon_{ab}{}^{cd}F_{cd}. 
\end{equation}
It then follows that
\begin{align}\lbeq{Fdecomp}
  F_{ab} &= 2E_{[a}\mu_{b]} - \epsilon_{ab}{}^{cd}B_c\mu_d \\
  *F_{ab} &= 2B_{[a}\mu_{b]} + \epsilon_{ab}{}^{cd}E_c\mu_d.
\end{align}
Assuming that $F_{ab}$ is stationary as well as source-free so that
$*F_{ab}$ is closed, then both $E_a$ and $B_a$ are gradients of
stationary scalars. Hence we introduce electric and magnetic
potentials $v$ and $u$ according to
\begin{equation}\lbeq{Fscalars}
  E_a = -\nabla_a v, \quad B_a = -\nabla_a u.
\end{equation}
Moreover, with the electromagnetic field being the only matter source,
a twist scalar $\chi$, likewise stationary, can be introduced
by\footnote{We use units such that the four-dimensional
Einstein-Maxwell equations read $G_{ab}=2T_{ab}$ where
$T_{ab}=F_a{}^cF_{bc}-\tsfrac14F^{cd}F_{cd}g_{ab}$.}
\begin{equation}\lbeq{twist}
  \nabla_{\!a}\chi = \epsilon_{abcd}\,\xi^b\nabla^d\xi^c + 2(u\nabla_{\!a}v-v\nabla_{\!a}u). 
\end{equation}
The tensor $h_{ab}$ can be viewed as a metric on the three-dimensional
manifold $\Sigma$ of Killing orbits and we shall use that the full 4D
Einstein-Maxwell equations are now equivalent to the 3D equations
obtained from the action
\begin{equation}\lbeq{action}
  S = \int d^3x\sqrt{h}\left[{}^{(3)}R - 2h^{ab}\gamma_{AB}D_a X^A D_b X^B\right],
\end{equation}
where ${}^{(3)}R$, $h^{ab}$ and $D_a$ are the Ricci scalar, inverse
and Levi-Civita connection of $h_{ab}$. This action describes
three-dimensional Euclidean gravity coupled to a certain sigma
model. More precisely, the effective matter part of the action is that
of a wave map from $\Sigma$ to a four-dimensional target space with
metric
\begin{equation}
  d\sigma^2 = \gamma_{AB}dX^A dX^B = \frac{\frac12 d\mathcal{E} d\bar{\mathcal{E}} + \psi\,d\bar\psi d\mathcal{E} + \bar\psi\,d\psi d\bar{\mathcal{E}} - (\mathcal{E}+\bar{\mathcal{E}})d\psi d\bar\psi}{2\left[\tsfrac12(\mathcal{E}+\bar{\mathcal{E}}) + \psi\bar\psi\right]^2},
\end{equation}
where $\mathcal{E}$ and $\psi$ are the complex Ernst potentials
\begin{equation}
  \mathcal{E} = f^2-v^2-u^2+i\chi, \quad \psi = v+iu.
\end{equation}
This is an Einstein metric, whose Ricci tensor obeys
\begin{equation}
  R_{AB} = -6\gamma_{AB}
\end{equation}
If we use as target space coordinates $X^A$ the real variables
$f,A,S,\chi$, where $A$ and $S$ are the amplitude and phase of $\psi$,
\ie
\begin{equation}
  \psi = Ae^{iS},
\end{equation}
the target metric takes the neat form
\begin{equation}
  d\sigma^2 = \frac1{f^2}(df^2-dA^2-A^2dS^2) + \frac{(d\chi+2A^2dS)^2}{4f^4},
\end{equation}
which makes it evident that it has signature $(-,-,+,+)$. Now, varying
the action \refeq{action} with respect to $h^{ab}$ gives
\begin{equation}\lbeq{R2harm}
  {}^{(3)}R_{ab} = 2\gamma_{AB}D_aX^A D_bX^B, 
\end{equation}
while variation with respect to the target space coordinates $X^A$
leads to the wave map equation
\begin{equation}\lbeq{wavemap}
  h^{ab}(D_a D_b X^C + \Gamma^C{}_{AB}D_aX^A D_bX^B) = 0,
\end{equation}
with $\Gamma^C{}_{AB}$ being the Christoffel symbols of
$\gamma_{AB}$. For our purposes it is important to note that if the
$X^A$ all depend solely on one free function $\omega$, then eqs.\
\refeq{R2harm} and \refeq{wavemap} become
\begin{align}
  &{}^{(3)}R_{ab} = 2\gamma_{AB}\frac{dX^A}{d\omega}\frac{dX^B}{d\omega}D_a\omega\,D_b\omega \\ \lbeq{geodesic}
  &\frac{dX^C}{d\omega}h^{ab}D_aD_b\omega + \left(\frac{d^2X^C}{d\omega^2}+\Gamma^C{}_{AB}\frac{dX^A}{d\omega}\frac{dX^B}{d\omega}\right)h^{ab}D_a\omega\,D_b\omega = 0.
\end{align}
Now, eq.\ \refeq{geodesic} tells us that the curve $X^A(\omega)$ is a
geodesic, since, as $h_{ab}$ is supposed to be a positive-definite
metric, it cannot happen that $h^{ab}D_a\omega\,D_b\omega$ vanishes
identically unless $\omega$ is constant. Using the freedom to
reparametrize the geodesic, we can assume that it is affinely
parametrized and arrive at the equations
\begin{align}\lbeq{epsilondef}
  &\epsilon := \gamma_{AB}\frac{dX^A}{d\omega}\frac{dX^B}{d\omega} = \mathrm{constant} \\
  &\frac{d^2X^C}{d\omega^2}+\Gamma^C{}_{AB}\frac{dX^A}{d\omega}\frac{dX^B}{d\omega} = 0 \\ \lbeq{EinKG}
  &{}^{(3)}R_{ab}=2\epsilon D_a\omega\,D_b\omega \\ \lbeq{laplace}
  &h^{ab}D_aD_b\,\omega = 0.
\end{align}
In this paper we focus exclusively on this class of solutions, which
we shall refer to as geodesic solutions.
\subsection{Axisymmetry and Weyl coordinates}
Let us now assume that the stationary spacetime is also axisymmetric
with an axisymmetry generator $\eta^a$ that commutes with the timelike
Killing vector $\xi^a$. We can then introduce coordinates
$(t,\rho,z,\phi)$ such that $\xi^a=(\partial/\partial t)^a$,
$\eta^a=(\partial/\partial\phi)^a$ while the three-metric $h_{ab}$ and
one-form $\mu_a$ take the forms
\begin{align}
  &dl^2 = e^{2k}(d\rho^2+dz^2) + W^2 d\phi^2 \\
  &\pmb{\mu} = dt+\Omega\,d\phi.
\end{align}
The metric functions $f$, $k$, $W$ and $\Omega$ obviously depend on
$\rho$ and $z$ only. If we assume that $\omega$ also depends solely on
$\rho$ and $z$ (although it could have a linear dependence on $\phi$
with constant coefficient, a possibility which we do not consider
here), the $\phi\phi$-component of eq.\ \refeq{EinKG} becomes
\begin{equation}
  W_{,\rho\rho}+W_{,zz} = 0,
\end{equation}
making it possible to choose $\rho$ and $z$ such that $W=\rho$, a
choice well-known as Weyl (canonical) coordinates. The great advantage
of being able to use these coordinates is that the unknown metric
function $k$ does not enter the Laplace equation \refeq{laplace} (nor
the more general wave map equation \refeq{wavemap}), which will thus
be identical to the Laplace equation in flat space, expressed in
cylindrical coordinates, for an unknown function that is independent
of $\phi$;
\begin{equation}\lbeq{flatlaplace}
  \omega_{,\rho\rho}+\rho^{-1}\omega_{,\rho}+\omega_{,zz} = 0.
\end{equation}
With the use of this equation, the remaining components of eq.\
\refeq{EinKG} become
\begin{equation}\lbeq{keq}
\begin{split}
  &k_{,\rho} = \epsilon\rho\left[(\omega_{,\rho})^2-(\omega_{,z})^2\right] \\
  &k_{,z} = 2\epsilon\rho\,\omega_{,\rho}\,\omega_{,z},
\end{split}
\end{equation}
while eq.\ \refeq{twist} gives
\begin{equation}\lbeq{Omegaeq}
\begin{split}
  &\Omega_{,\rho} = 4p_\chi\,\rho\,\omega_{,z} \\
  &\Omega_{,z} = -4p_\chi\,\rho\,\omega_{,\rho},
\end{split}
\end{equation}
with $p_\chi$ being a constant equal to the conserved geodesic
momentum associated with the cyclic coordinate $\chi$ of the target
space metric. Explicitly,
\begin{equation}\lbeq{pchi}
  p_\chi = \frac{\displaystyle\frac{d\chi}{d\omega}+2A^2\frac{dS}{d\omega}}{4f^4}.
\end{equation}
A recipe for finding a solution of the type considered here is thus to
choose an appropriate geodesic of the target space and an appropriate
solution to the flat space Laplace equation \refeq{flatlaplace}. The
geodesic then gives the four target space functions as functions of
$\omega$, including the metric component $f$ and the potentials $v$
and $u$ which completely determine the electromagnetic field via eqs.\ 
\refeq{Fdecomp} and \refeq{Fscalars}. What then remains is to
integrate the two pairs of equations \refeq{keq} and \refeq{Omegaeq}
to determine the remaining spacetime metric components $k$ and
$\Omega$. Note that eq.\ \refeq{flatlaplace} is the integrability
condition for both pairs of equations (when $\epsilon\neq 0\neq
p_\chi$).

\section{The Reissner-Nordstr\"om-NUT solution}%
Using Schwarzschild type coordinates, the Reissner-Nordstr\"om-NUT
(RNN) solution with mass $M$, electric charge $Q$ and NUT
(gravitational monopole) charge $l$ is
\begin{align}\lbeq{rnn}
  ds^2 = -f^2(dt+\Omega d\phi)^2+f^{-2}dr^2+(r^2+l^2)(d\theta^2+\sin^2\!\theta d\phi^2), \\ \lbeq{Frnn}
  \pmb{F} = Q\,\frac{r^2-l^2}{(r^2+l^2)^2}(dt+\Omega d\phi)\wedge dr + \frac{2Qlr}{r^2+l^2}\sin\theta\,d\theta\wedge d\phi. 
\end{align}
where
\begin{align}
  &f^2 = \frac{r^2-2Mr+Q^2-l^2}{r^2+l^2} = \frac{(r-r_+)(r-r_-)}{r^2+l^2}
 \\ \lbeq{Omegaschw}
  &\Omega = 2l\cos\theta + \Omega'.
\end{align}
Here we have defined
\begin{equation}
  r_\pm = M \pm \Delta, \quad \Delta = \sqrt{M^2+l^2-Q^2},
\end{equation}
while $\Omega'$ is a constant which should be set to $-2l$ ($2l$) to
make the half-axis $\theta=0$ ($\theta=\pi$) explicitly regular,
leaving the other half-axis -- the Misner string -- singular, as
$d\phi$ is not a well-behaved one-form at $\theta=0,\pi$.  However, it
is well-known that both half-axes can be made regular, since changing
$\Omega'$ from $-2l$ to $2l$ can be mimicked by changing the time
coordinate from $t$ to $t' = t - 4l\phi$. The price to pay is closed
timelike curves since it requires that $t$ and $t'$ should both be
periodic with period $8\pi l$.

The three-metric $h_{ab}$ for this solution is seen to be
\begin{equation}
  dl^2 = dr^2 + (r^2-2Mr+Q^2-l^2)d\Omega^2 = dr^2 + (r-r_+)(r-r_-)d\Omega^2
\end{equation}
and the three remaining target space scalars are given by
\begin{align} 
  &\chi = \frac{2l(r-M)}{r^2+l^2} \\
  &A = \frac{Q}{\sqrt{r^2+l^2}} \\
  &S = -\arctan{\!\left(\frac{l}{r}\right)},
\end{align}
Since all the $X^A$ depend on $r$ only, this is a geodesic
solution. In the non-extremal case $\Delta>0$, the geodesic is
spacelike ($\epsilon>0$) and explicitly given by
\begin{align}\lbeq{Fgeod}
  &f = \frac{\Delta}{\sqrt{(M\sinh{\omegaAL}-\Delta\cosh{\omegaAL})^2+l^2\sinh^2\!{\omegaAL}}} 
  = \frac{r_+-r_-}{\sqrt{(r_- e^{\omegaAL}-r_+e^{-{\omegaAL}})^2+l^2(e^{\omegaAL}-e^{-{\omegaAL}})^2}} \\
  &\chi = -\frac{l\Delta\sinh{2{\omegaAL}}}{(M\sinh{\omegaAL}-\Delta\cosh{\omegaAL})^2+l^2\sinh^2\!{\omegaAL}} 
  = -\frac{2l\Delta(e^{2{\omegaAL}}-e^{-2{\omegaAL}})}{(r_- e^{\omegaAL}- r_+ e^{-{\omegaAL}})^2+l^2(e^{\omegaAL}-e^{-{\omegaAL}})^2} \\
  &A = -\frac{Q\,\sinh{\omegaAL}}{\sqrt{(M\sinh{\omegaAL}-\Delta\cosh{\omegaAL})^2+l^2\sinh^2\!{\omegaAL}}} 
  = -\frac{Q\,(e^{\omegaAL}-e^{-{\omegaAL}})}{\sqrt{(r_- e^{\omegaAL}- r_+ e^{-{\omegaAL}})^2+l^2(e^{\omegaAL}-e^{-{\omegaAL}})^2}} \\
  &S = -\arctan{\!\left(\frac{l\sinh{\omegaAL}}{M\sinh{\omegaAL}-\Delta\cosh{\omegaAL}}\right)} 
  = -\arctan{\!\left[\frac{l\,(e^{\omegaAL}-e^{-{\omegaAL}})}{r_-e^{\omegaAL}-r_+e^{-{\omegaAL}}}\right]},
\end{align}
where $\omegaAL$ is the arclength parameter for the geodesic, and here
the function that satisfies the Laplace equation \refeq{laplace},
namely
\begin{equation}
  \omegaAL = -\arccoth{\!\left(\frac{r-M}{\Delta}\right)} =
  \frac12\ln{\!\left(\frac{r-r_+}{r-r_-}\right)}.
\end{equation}
However, we shall instead think of the geodesic as parametrized by the
rescaled affine parameter
\begin{equation}
  \omega = \frac{M}{\Delta}\,\omegaAL,
\end{equation}
which implies that the norm of the geodesic tangent vector has norm
$\epsilon = \Delta/M$. The reason for this is that we will then
straightforwardly be able to treat the extremal case -- which
corresponds to a null geodesic ($\epsilon = 0$) -- as the $\Delta\rarr
0$ limit of the non-extremal case. Indeed, taking this limit for the
above formulae, we find that the spacelike geodesic goes over into a
lightlike one;
\begin{align}\lbeq{Fextr}
  &f = \frac{M}{\sqrt{M^2(1-\omega)^2+l^2\omega^2}} \\
  &\chi = -\frac{2lM\omega}{M^2(1-\omega)^2+l^2\omega^2} \\
  &A = -\frac{Q\omega}{\sqrt{M^2(1-\omega)^2+l^2\omega^2}} \\
  &S = \arctan{\!\left[\frac{l\omega}{M(1-\omega)}\right]},
\end{align}
while $\omega$ goes over into
\begin{equation}
  \omega = -\frac{M}{r-M}.
\end{equation}
This limiting procedure would not have worked, had we used
$\omegaAL$ as the affine parameter. Moreover, $\omega$ has
the large $r$ asymptotic behaviour
\begin{equation}
  \omega = -\frac{M}{r} + O(r^{-2}),
\end{equation}
which means that it is $\omega$, rather than $\omegaAL$,
that corresponds to a Newtonian gravitational potential. However,
although we will think of the geodesic as parametrized by $\omega$, we
will often work with $\omegaAL$ in subsequent calculations,
since the factor $M/\Delta$ would otherwise often appear merely as an
annoying appendage. 

For future reference, we here finally calculate the conserved geodesic
momentum $p_\chi$ according to the formula \refeq{pchi}, to find the
simple relation
\begin{equation}\lbeq{pchicalc}
  p_\chi = -\ds\frac{l}{2M}.  
\end{equation}

\subsection{Charges}
A special and somewhat surprising feature of the RNN solution is that
although no sources are included for the electromagnetic field, the
result of calculating the electric charge $Q_\mathrm{el}$ by
integrating $*F_{ab}$ over a two-sphere of constant $t$ and $r$ and
dividing by $4\pi$ is not what one would naively expect from Stoke's
theorem, \ie the result is not independent of the choice of two-sphere
but depends on the radius;
\begin{equation}
  Q_\mathrm{el} = Q\,\frac{r^2-l^2}{r^2+l^2}.
\end{equation}
We also calculate the magnetic charge $Q_\mathrm{mag}$ by integrating
$F_{ab}$ over the same sphere to find
\begin{equation}
  Q_\mathrm{mag} = \frac{2Qlr}{r^2+l^2}.
\end{equation}
Obviously we have $Q_\mathrm{el} = Q$, $Q_\mathrm{mag}=0$
asymptotically as $r\rarr\infty$, so at infinity the solution is
purely electric with charge $Q$. It is worth noting that the relation
\begin{equation}
  Q_\mathrm{el}^{\,2} + Q_\mathrm{mag}^{\,2} = Q^2 
\end{equation}
holds as an identity, for all $r$. The reason why it is possible that
electric and magnetic charges depend on $r$ is that the
gravitomagnetic vector potential $\Omega d\phi$ enters not only the
metric but also the electromagnetic field strength \refeq{Frnn}. Thus
the latter is singular on the Misner string, \ie on $\theta = 0$ or
$\theta = \pi$ (or both) depending on how $\Omega'$ is chosen. As the
Misner string goes through every two-sphere that we integrate over, we
cannot expect Stoke's theorem to imply that the charges should be
independent of the choice of sphere. Interestingly, if we refrain from
removing the singularity by introducing periodic time, the Misner
string must be thought of as a concrete physical object as it carries
electric and magnetic charge density.

\subsection{Transformation to Weyl coordinates}
For the RNN solution, the transformation from Schwarzschild type
coordinates to Weyl coordinates reads
\begin{align}
  &\rho = \sqrt{(r-M)^2-\Delta^2}\sin\theta \\
  &z = (r-M)\cos\theta,
\end{align}
with inverse transformation
\begin{align}
  &r = \lambda + M \\
  &\cos\theta = \frac{z}{\lambda},
\end{align}
where
\begin{equation}
  \lambda = \frac12(\lambda_+ + \lambda_-), \quad \lambda_\pm = \sqrt{\rho^2+(z\pm\Delta)^2}. 
\end{equation}
The functions $\omega$, $k$ and $\Omega$ are now given by
\begin{align}\lbeq{omegaorig}
  &\omega = \frac{M}{2\Delta}\ln{\!\left(\ds\frac{\lambda-\Delta}{\lambda+\Delta}\right)} \\
  &k = \frac12\ln{\!\left(\frac{\lambda^2-\Delta^2}{\lambda^2-\eta^2}\right)}, \quad \eta = \frac12(\lambda_+-\lambda_-) \\ \lbeq{Omegaweyl}
  &\Omega = \frac{2lz}{\lambda} + \Omega'.
\end{align}
In the non-extremal case $\Delta>0$, the function $\omega$ is the
potential of an infinitely thin rod located at $\rho=0$,
$\abs{z}\leq\Delta$ and having a line density $M/(2\Delta)$ per unit
length, which means that the line density can be interpreted as the
mass line density.  Note that in the extremal case $\Delta = 0$, the
function $k$ vanishes identically leaving the three-metric $h_{ab}$
flat.  Moreover $\lambda$ reduces to $\sqrt{\rho^2+z^2}$ and $\omega$
to the standard monopole solution $-M/\lambda$.

\section{Periodic analogue of the Reissner-Nordstr\"om-NUT solution}
We will now construct compactified versions of the RNN solution, which
generalize the compactified Schwarzschild solution discussed in
\cite{Myers,Korotkin1,Korotkin2,Frolov}. The 
approach we take can be summerized as follows:
\begin{itemize}
\item For every member of the RNN family of solutions, compactify the
  function $\omega$ by using canonical Weyl coordinates and taking the
  $z$-periodic analogue of the original solution to the Laplace
  equation \refeq{flatlaplace}. This is straightforward due to the
  linearity of the latter.
  \item Insert $\omega$ into the
  \emph{same} geodesic that defines the original RNN solution.  
  \item
  Integrate eqs.\ \refeq{keq} and \refeq{Omegaeq} to find the
  remaining metric functions $k$ and $\Omega$.
\end{itemize}   
Before we start, some general remarks about the first of these steps
are in order. As proved by Korotkin and Nicolai
\cite{Korotkin1,Korotkin2}, if 
$\omegaorig(\rho,z)$ is a solution to the Laplace equation
with the asymptotic behaviour
\begin{equation}
  \omegaorig(\rho,z) = -\frac{M}{\tilde{r}} + O(\tilde{r}^{-2}) \quad \mbox{as
  $\tilde{r}\rarr\infty$},
\end{equation}
where $\tilde{r}=\sqrt{\rho^2+z^2}$ and $M$ is some constant (in our
case the mass), then the series
\begin{equation}\lbeq{omegaperiodic}
  \omega(\rho,z) = \sum_{n=-\infty}^\infty\left[\,\omegaorig(\rho,z+nL) + a_n\,\right], \qquad a_n = \left\{ \begin{array}{ll} \ds\frac{M}{L\abs{n}}  & \textrm{if $n\neq 0$} \\ 
                                0 & \textrm{if $n=0$} 
       \end{array}\right. ,
\end{equation}
is convergent for any $(\rho,z)$ such that $(\rho,z+nL)$ does not
coincide with a singular point of $\omegaorig(\rho,z)$ for
any integer $n$. The resulting function $\omega$ then obviously
defines a $z$-periodic solution to the Laplace equation
\refeq{flatlaplace} with period $L$. The constants $a_n$ are essential as
they make the series converge, but one cannot say that they are
uniquely determined as one could add to $a_n$ any other $n$-dependent
constant $b_n$ that falls off faster than $\abs{n}^{-1}$ as
$\abs{n}\rarr\infty$ so that
\begin{equation}
  B := \sum_{n=-\infty}^\infty b_n 
\end{equation}
is a finite constant. In other words, the $z$-periodic analogue of the
function $\omegaorig(\rho,z)$ is \apriori only defined up to
an additive constant $B$. Korotkin and Nicolai fix this constant by
taking eq.\ \refeq{omegaperiodic} as it stands ($B=0$) to define
$\omega(\rho,z)$, but this choice has not been made by the other
workers that have studied the compactified Schwarzschild black hole;
while Myers\cite{Myers} makes an explicitly different
choice of $a_n$ leading to a certain non-zero $B$, Frolov and
Frolov\cite{Frolov} instead use a Green's function method to
compactify which in effect corresponds to a third choice of $B$. Now,
since the function $\omega(\rho,z)$ is periodic in $z$, it does not
make sense to study its behaviour for large $\tilde{r}$, but for large
$\rho$ it has the asymptotic behaviour
\begin{equation}
  \omega = \frac{2M}{L}\,\ln\rho + O(1). 
\end{equation}
If $\omega$ would tend to a constant in this limit, it would be
natural to make that constant vanish by a suitable choice of $B$.
However, since $\omega$ instead diverges logarithmically (unless $M =
0$), we see no physical motivation for fixing $B$ at any particular
value and hence we will keep it as a free parameter.

We will now implement the compactification scheme as outlined above.
We thus take $\omegaorig$ to be the function given by eq.\ 
\refeq{omegaorig}. According to the above, its periodic analogue is
\begin{equation}
  \omega = \sum_{n=-\infty}^\infty
  \left[\frac{M}{2\Delta}\,\ln{\!\!\left(\frac{\lambda_n-\Delta}{\lambda_n+\Delta}\right)}
  + a_n\right] + B,
\end{equation}
where
\begin{align}
  &\lambda_n = \frac12\left[\sqrt{\rho^2+(z_n+\Delta)^2}+\sqrt{\rho^2+(z_n-\Delta)^2}\right], \quad z_n = z-nL \\
  &a_n = \left\{ \begin{array}{ll} \ds\frac{M}{L\abs{n}}, & \textrm{if $n\neq 0$} \\
                                    0 & \textrm{if $n=0$}. \\
        \end{array} \right. 
\end{align}
We take $z$ to have the range $z\in [-L/2,L/2]$ with the end points of
the interval identified. The horizon, where $\omega$ diverges to minus
infinity, will thus be located at $\rho=0$, $\abs{z}\leq\Delta$, just
as in the non-compactified case. Since we shall not consider the case
when the horizon overlaps itself, we require that $\Delta\leq L/2$. It
will prove useful to introduce the dimensionless variable
\begin{equation}
  \beta = \frac{2\Delta}{L},
\end{equation}
having the range $\beta\in[0,1]$. For convenience, we will, for the
time being, work with $\omegaAL$ and use the dimensionless
coordinates
\begin{equation}
  x=\frac{\rho}{\Delta}, \quad y=\frac{z}{\Delta}.
\end{equation}
Clearly $\omegaAL$ is then given by 
\begin{equation}\lbeq{omegaALsum}
  \omegaAL = \sum_{n=-\infty}^\infty
  \left[\frac12\ln{\!\!\left(\frac{\lambda'_n-1}{\lambda'_n+1}\right)}
  + \bar{a}_n\right] + \bar{B},
\end{equation}
where
\begin{align}
  &\lambda'_n = \frac12\left[\sqrt{x^2+(y_n+1)^2}+\sqrt{x^2+(y_n-1)^2}\right], \quad y_n = y-2n\beta^{-1} \\
  &\bar{a}_n = \frac{\Delta}{M}a_n = \left\{ \begin{array}{ll} \ds\frac{\beta}{2\abs{n}}, & \textrm{if $n\neq 0$} \\
                                    0 & \textrm{if $n=0$}, \\
        \end{array} \right. \\
  &\bar{B} = \frac{\Delta}{M}\,B.
\end{align}
We shall now explicitly evaluate $\omegaAL$ close to the
symmetry axis $x=0$. As the behaviour on the horizon section
$\abs{y}\leq 1$ of the axis is different from the off-horizon section
$1<\abs{y}\leq\beta^{-1}$, these two sections will have to be treated
differently, but in both cases we shall express the result in terms of
the function
\begin{equation}
  \varphi_\beta(\xi) = \frac{\beta}{2\pi}\sin{\!\parfrac{\pi\beta\,\xi}2}\,\Gamma{\parfrac{\beta\,\xi}2}^{\!\!2},
\end{equation}
having the properties
\begin{align}
  &\lim_{\beta\rarr 0}\varphi_\beta(\xi) = \frac1{\xi} \\
  &\varphi_\beta(\xi)\varphi_\beta(-\xi) = -\frac1{\xi^2}. 
\end{align}
Moreover, for reasons that will become clear, we shall replace the
free constant $\bar{B}$ by a free constant $\bar{u}$ according to 
\begin{equation}
  \bar{B} = \bar{u} - \beta\gamma - \frac12\ln{[2\varphi_\beta(2)]},
\end{equation}
where $\gamma$ is the Euler constant. Now, for small $x$ and
$\abs{y}\leq 1$, we find that $\omegaAL$ is given by
\begin{equation}
  e^{2\omegaAL} = \frac{e^{2\bar{u}}}{4H(y)}\,x^2 + O(x^4),  
\end{equation}
where $H(y)$ is the function
\begin{equation}
  H(y) = \frac{2\varphi_\beta(2)}{\varphi_\beta(1+y)\varphi_\beta(1-y)}. 
\end{equation}
For $1<\abs{y}\leq\beta^{-1}$,
on the other hand, one finds that
\begin{equation}
  e^{2\omegaAL} = \frac{e^{2\bar{u}}}{2\varphi_\beta(2)} 
  \frac{\varphi_\beta(\abs{y}+1)}{\varphi_\beta(\abs{y}-1)} + O(x^2). 
\end{equation}
This is all that is needed to generate the whole solution. Indeed, for
both axis sectors, $\omegaAL$ is a solution to the Laplace equation
which can be expanded in terms of a constant $C$ and a function $A(y)$
as
\begin{equation}\lbeq{omegaexpx}
  \omegaAL = C\ln{x} + \sum_{k=0}^\infty\frac{(-1)^k}{(2^k k!)^2}\frac{d^{2k}A(y)}{dy^{2k}}\,x^{2k}. 
\end{equation}
Inserting this expansion into the right hand sides of eqs.\ 
\refeq{keq} and \refeq{Omegaeq} (using eq.\ \refeq{pchicalc} for the
latter), we find that $k$ and $\Omega$ are determined up to additive
constants $k'$ and $\Omega'$;
\begin{align}\lbeq{kexpx}
  &k = k' + C^2\ln{x} + 2CA(y) -\frac12\left[CA''(y)+A'(y)^2\right]x^2 + O(x^4) \\ \lbeq{Omegaexpx}
  &\Omega = \Omega' + 2l\left[Cy + \sum_{k=1}^\infty\frac{2k\,(-1)^k}{(2^kk!)^2}\frac{d^{2k-1}A(y)}{dy^{2k-1}}\,x^{2k}\right],
\end{align}
where we refrain from giving the whole series for $k$ since it depends
nonlinearly on $\omegaAL$, although it would not be difficult to write
down at least a few more terms. The value of $k'$ has to be determined
from the requirement that the axis be regular. Starting with the
off-horizon section, regularity means absence of a conical singularity
which is in turn means that $k$ should vanish as $x\rarr 0$, at least
if we assume that $\Omega$ vanishes in that limit, which can always be
achieved as we shall see below.  Thus, since the constant $C$ is zero
in this case, we directly obtain $k' = 0$.  For the horizon section we
instead have $C=1$. In this case we may determine $k'$ by inspecting
the horizon metric, which reads
\begin{equation}
  ds_\mathrm{H}^2 = R^2\left[\frac{e^{2(k'+\bar{u})}}{16}\frac{dy^2}{H(y)}+H(y)d\phi^2\right],
\end{equation}
while $R$ is the constant
\begin{equation}
  R = \sqrt{r_+^2+l^2}\,e^{-\bar{u}}
\end{equation}
Now, since
\begin{equation} 
  H'(\pm 1) = \mp 2,
\end{equation}
it follows that for the horizon's polar points $y=\pm 1$ to be regular, we must set
\begin{equation}
  k' = -\bar{u} + \ln{4}. 
\end{equation}
When we are on the horizon $\abs{y}\leq 1$, the constant $\Omega'$ in
eq.\ \refeq{Omegaexpx} precisely corresponds to the constant
$\Omega'$ in eqs.\ \refeq{Omegaschw} and \refeq{Omegaweyl} and should
be set to $-2l$ ($2l$) to make $\Omega$ vanish at the horizon pole
$y=1$ ($y=-1$). Continuity at $y=\pm 1$ then requires us to set
$\Omega' = 0$ for $y>1$ ($y<1$) and $\Omega'=-4l$ ($\Omega'=4l$) for
$y<1$ ($y>1$). Again, these two choices do not correspond to different
physics, but are related by a change of time $t\rarr t' = t-4l\phi$
which results in time periodicity with the period $8\pi l$. However,
note that unlike the non-periodic case, both choices imply that
$\Omega$ has a jump discontinuity where $y=\beta^{-1}$ is periodically
identified with $y=-\beta^{-1}$. This discontinuity is of course not
of a physical nature either, since we can shift it to any other
position on the $y$-axis. For instance, we may put the discontinuity
somewhere on the horizon and obtain a completely regular off-horizon
section on which $\Omega$ is everywhere zero.



\subsection{Fourier expansions}\label{sec:fourier}
\begin{figure}[ht]
  \centering \begin{minipage}[t]{0.8\linewidth} \centering
  \includegraphics[width=0.95\textwidth, angle=0]{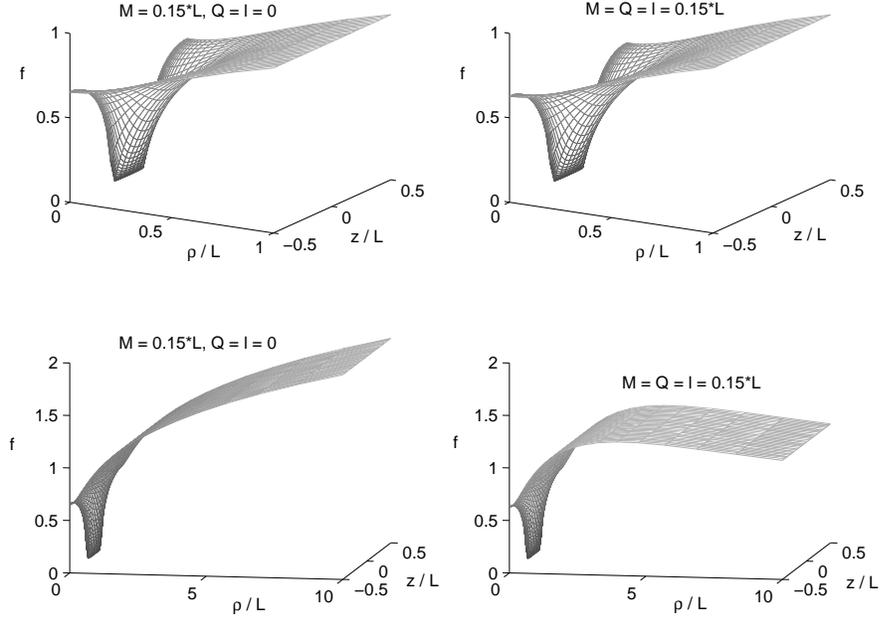}
  \caption{The metric function $f = \sqrt{-g_{tt}}$, plotted for $M = 0.15L$, 
    $Q = l = 0$ as well as for $M = Q = l = 0.15L$, with $\bar{u}=0$
    in both cases. As the two top plots show, the behaviour of $f$ for
    the two cases is very similar close to the axis $\rho = 0$. On the
    other hand $f$ behaves qualitatively different when a larger range
    of $\rho$ is included, as indicated by the plots below; in the
    first case $f$ increases monotonically with $\rho$ towards
    infinity, while in the second case $f$ reaches a maximum and
    subsequently decreases towards zero. The latter behaviour is
    generic as long as $l\neq 0$, regardless of $Q$, but with $l = 0
    \neq Q$, the maximum is replaced by a divergence, as discussed in
    subsection \ref{sec:large}.}
  \label{fig:gravpot}
  \end{minipage}
\end{figure}

In principle, the full RNN solution is given as the $x=0$ expansions
\refeq{omegaexpx} - \refeq{Omegaexpx}, but since these series converge
very slowly for large $x$ and are quite useless when it comes to
determining the asymptotic behaviour as $x\rarr\infty$, we shall now
follow Frolov and Frolov and represent the solution in terms of
fourier series, thus making explicit use of the periodicity of the
coordiante $y$. Still using the rescaled coordinates $(x,y)$, we
assume fourier expansions for the functions $\omegaAL$ and $k$ of the
forms
\begin{align}
  &\omegaAL = \sum_{q=0}^\infty\omegaAL_q(x)\cos{\!(\pi q\beta y)} \\
  &k = \sum_{q=0}^\infty k_q(x)\cos{\!(\pi q\beta y)} 
\end{align}
Since $\omegaAL$ is a solution to the Laplace equation
\refeq{flatlaplace}, it follows that
\begin{equation}
  \omegaAL_q''(x) + x^{-1}\omegaAL_q'(x) - (\pi q\beta)^2\omegaAL_q(x) = 0. 
\end{equation}
Disqualifying solutions that diverges exponentially as $x\rarr\infty$,
we obtain that $\omegaAL_0(x)$ is a linear function of $\ln{x}$ while,
for $q>0$,
\begin{equation}
  \omegaAL_q(x) \propto K_0(\pi q\beta x), 
\end{equation}
where $K_0$ is the modified Bessel function of the second kind. For
small $x$, we have
\begin{equation}
  -K_0{(\pi q\beta x)} = \gamma + \ln{\!\parfrac{\pi q\beta x}2} + O(x^2).
\end{equation}
We can now use that we already know $\omegaAL$ for small $x$ to find
that 
\begin{align}
  &\omegaAL_0(x) = \bar{u} - \frac12\ln{[2\varphi_\beta(2)]} + \beta\,\ln{\!\parfrac{\beta\,x}4} \\
  &\omegaAL_q(x) = -\frac2{\pi q}\,\sin{\!(\pi q\beta)}\,K_0{(\pi q\beta x)} \quad \mbox{for $q>0$}. 
\end{align}
We use this fourier expansion for $\omegaAL$ to plot the metric
function $f$ for two choices of the parameters $M$, $Q$ and $l$
(figure \ref{fig:gravpot}).

Due to the nonlinearity of eqs.\ \refeq{keq} which
governs $k$, we shall not attempt to give its full fourier series.
However, since all $k_q(x)$ with $q>0$ decay exponentially as
$x\rarr\infty$, it will suffice to determine $k_0(x)$ to obtain the
asymptotic behaviour in that limit. Now, the zeroth order fourier term
of the first of eqs.\ \refeq{keq} reads
\begin{equation}
  k_0'(x) = x\left[(\omegaAL_{,x})^2-(\omegaAL_{,z})^2\right]_0 
  = x\left\{ \omegaAL_0'(x)^2 + \frac12\sum_{q=1}^\infty
  \left[\omegaAL_q'(x)^2-(\pi q\beta)^2\omegaAL_q(x)^2\right]\right\} 
\end{equation}
Inserting the determined expressions for the $\omegaAL_q(x)$ gives
\begin{equation}
  k_0'(x) = \beta^2 x\left\{\frac1{x^2} + 2\sum_{q=1}^\infty\sin^2{\!(\pi q\beta)}\left[K_1{(\pi q\beta x)}^2 - K_0{(\pi q\beta x)}^2\right]\right\},
\end{equation}
which can be integrated to yield
\begin{equation}\lbeq{k0}
  k_0(x) = k_* + \beta^2\left\{\ln{x} + x^2\sum_{q=1}^\infty\sin^2{\!(\pi q\beta)}\left[2K_1{(\pi q\beta x)}^2 - K_0{(\pi q\beta x)}K_2{(\pi q\beta x)}- K_0{(\pi q\beta x)}^2\right] \right\},
\end{equation}
where $k_*$ is some constant yet to be determined. Using now that 
\begin{equation}
  x^2\left[2K_1{(\pi k\beta x)}^2 - K_0{(\pi k\beta x)}K_2{(\pi k\beta x)}- K_0{(\pi k\beta x)}^2\right] = \frac2{(\pi k\beta)^2}\left[1+\gamma+\ln{\!\parfrac{\pi k\beta x}2}\right] + O(x^2),
\end{equation}
we can compare $k_0(x)$ as given by eq. \refeq{k0} to the zeroth order
fourier term of the exact expression for $k$ we have near $x=0$ to
find, after some manipulations, that $k_*$ is the $\beta$-dependent
constant
\begin{equation}
  k_* = \beta^2\left[1+\ln{\!\parfrac{\beta}4}\right] - \beta\ln{\![\varphi_\beta(2)]} + \int_0^\beta\!\ln{\![\varphi_{\beta'}(2)]}d\beta'. 
\end{equation}
Finally, from eqs.\ \refeq{Omegaeq} and \refeq{pchicalc} we obtain
that $\Omega$ is given by the following ``quasi-fourier'' series:
\begin{align}
 \Omega = \Omega' + 2l\beta y + 4l\sum_{k=1}^\infty\frac1{\pi k}\sin{\!\left(\pi k\beta\right)}\,xK_1{\left(\pi k\beta x\right)}\sin{\!\left(\pi k\beta y\right)},
\end{align}
where $\Omega'$ is the same constant that appears in eq.\ 
\refeq{Omegaexpx} for the case $\abs{y}\leq 1$.

\subsection{Properties of the event horizon}
As noted above, the (outer) event horizon has the geometry
\begin{equation}
  ds_\mathrm{H}^2 = \mathcal{R}^2\left[\frac{dy^2}{H(y)}+H(y)d\phi^2\right], \quad 
  \mathcal{R} = e^{-\bar{u}}\sqrt{r_+^2+l^2}. 
\end{equation}
It directly follows that the horizon area and Gaussian curvature (half
the Ricci scalar) are given by the simple formulae
\begin{align}
  &\mathcal{A} = 4\pi\mathcal{R}^2 \\
  &K = -\frac12 H''(y)\mathcal{R}^{-2}. 
\end{align}
It is worth noting that $H(y)$ behaves for small $\beta$ as
\begin{equation}\lbeq{sphere}
  H(y) = 1-y^2 - \frac12\zeta(3)(1-y^2)^2\beta^3 + O(\beta^5),
\end{equation}
while for $\beta$ close to unity it behaves as
\begin{equation}\lbeq{gar}
  H(y) = 4(1-\beta) + 4\left[\Psi{\parfrac{1+y}2}+\Psi{\parfrac{1-y}2}+2\gamma+1\right](1-\beta)^2 + O((1-\beta)^3), 
\end{equation}
where $\Psi$ is the digamma function (the logarithmic derivative of
the gamma function). Clearly, for $\beta=0$, the metric is an exact
two-sphere of radius $\mathcal{R}$. Note that $\beta\rarr 0$
corresponds to two independent and physically different limits, namely
the extremal limit $\Delta\rarr 0$ as well as the limit of infinite
coordinate diameter $L\rarr\infty$. This is natural since both limits,
makes the \emph{proper} distance between the poles of the black hole
infinite, keeping the black hole from being distorted ``by itself'',
\ie by the gravitational field from its periodic copies. As $\beta$ is
increased, the sphere becomes deformed in a prolate manner. For small
$\beta$ the deformation is very small, however, as the first
correction term in eq.\ \refeq{sphere} enters at order $\beta^3$. As
$\beta$ is further increased the deformed sphere gets stretched out to
a long cigar, and for $\beta$ close to unity the geometry is almost
everywhere a flat cylinder of length $\mathcal{R}/\sqrt{1-\beta}$ and
circumference $4\pi\mathcal{R}\sqrt{1-\beta}$, except at the poles
$y=\pm 1$ where the curvature is very high (note that the mean
Gaussian curvature is always $\mathcal{R}^{-2}$). Up to a constant
conformal factor the geometry only depends on the parameter $\beta$
and hence we can refer the reader to Frolov and Frolov\cite{Frolov}
for a further discussion of its properties, as well as embedding
diagrams. Now, the horizon area $\mathcal{A}$ is clearly modified by
a factor $e^{-2\bar{u}}$ compared to the non-compactified case. This
is in agreement with the general treatment of distorted black holes
with (electric) charge as presented by Fairhurst and
Krishnan\cite{Fairhurst:2000xh}, based on Geroch and Hartle's
treatment of the case without charge\cite{Geroch}. Although it may not
have been obvious from the way we introduced it, the parameter
$\bar{u}$ (denoted the same way in \cite{Fairhurst:2000xh} but with an
unbarred $u$ in \cite{Geroch} and \cite{Frolov}) can be defined as
\begin{equation}
  \bar{u} = \delta\omegaAL|_{\rho = 0, z = \pm\Delta}, \quad 
  \delta\omegaAL = \omegaAL - \omegaALorig,  
\end{equation}
\ie $\bar{u}$ is the value of $\delta\omegaAL$ evaluated at either of
the poles of the event horizon, with $\delta\omegaAL$ being the
difference between the function $\omegaAL$ for the deformed black hole
(whether the deformation is due to compactification as in our case, or
to external matter) and the function $\omegaALorig$ for the undeformed
RNN solution with the same $M$, $Q$ and $l$. Since as $\omegaAL\rarr
-\infty$, we have
\begin{equation}
  f = \frac{2\Delta}{\sqrt{r_+^2+l^2}}\,e^{\omegaAL} + O(e^{3\omegaAL}), 
\end{equation}
it follows that $\bar{u}$ is closely related to the change in horizon
pole redshift factor that the deformation produces. In the
Schwarzschild case $Q = l = 0$, the relation is more direct since $f =
e^{\omegaAL}$ in that case. The term ``horizon redshift factor'' must
not be taken too literally, however, since it normally refers to an
asymptotically flat situation when the value of $f^{-1}$ gives the
redshift factor between the point of evaluation and infinity where
$f=1$. Anyhow, we shall refer to $\bar{u}$ as the \emph{redshift
  parameter}. In table \ref{table:ubar} we display how $\bar{u}$ has
in effect been fixed by previous workers that have studied the
compactified Schwarzschild solution. As we see no reason why any one
of these choices should be better than the other two, we prefer to
keep $\bar{u}$ as a free parameter. 

Other quantities that are interesting to calculate on the horizon are
the surface gravity $\kappa$ and Komar mass $M_\mathrm{Komar}$ with
respect to the Killing vector $\xi^a$, as well as the electric and
magnetic charges and potentials. We find that these quantities are
given by
\begin{align}
  &\kappa = \frac{\Delta}{\mathcal{R}^2}  \\
  &M_\mathrm{Komar} = \Delta \\
  &Q_\mathrm{el} = Q\frac{r_+^2-l^2}{r_+^2+l^2} \\
  &Q_\mathrm{mag} = \frac{2Qlr_+}{r_+^2+l^2} \\
  &v = -\frac{Qr_+}{r_+^2+l^2} \\
  &u = \frac{Ql}{r_+^2+l^2}
\end{align}
We note that the surface gravity is constant over the event horizon
and thus the zeroth law of thermodynamics holds. Moreover the electric
and magnetic potentials are constant as well. In fact the
electromagnetic charges and potentials take the exact same values as
in the noncompactified case. For a discussion of further thermodynamic
properties, we refer the reader to the general framwork for distorted
charged black holes\cite{Fairhurst:2000xh,Yazadjiev:2000by}.
Unfortunately, to our knowledge no such framework exists in the case
of nonzero NUT charge, but providing it here would be beyond the scope
of this paper.

\begin{table}
\begin{tabular}{|c|c|}
\hline
workers & $\bar{u}$ \\
\hline
Myers \cite{Myers} & $\frac12\ln{\![2\varphi_\beta(2)]} + \ln{\!\left[\frac{\Gamma(1-\beta/2)}{\Gamma(1+\beta/2)}\right]}$ \\
\hline
Korotkin and Nicolai \cite{Korotkin1,Korotkin2} & $\frac12\ln{\![2\varphi_\beta(2)]} + \beta\gamma$ \\
\hline
Frolov and Frolov \cite{Frolov} & $\frac12\ln{\![2\varphi_\beta(2)]} + \beta\ln{\!(4\pi)}$ \\
\hline
\end{tabular}
\caption{The redshift parameter $\bar{u}$ as used in previous studies of the compactified Schwarzschild solution.}
\label{table:ubar}
\end{table}
\subsection{Proper distance between black hole poles}
\begin{figure}[ht]
  \centering \begin{minipage}[t]{0.8\linewidth} \centering
  \includegraphics[width=0.95\textwidth, angle=0]{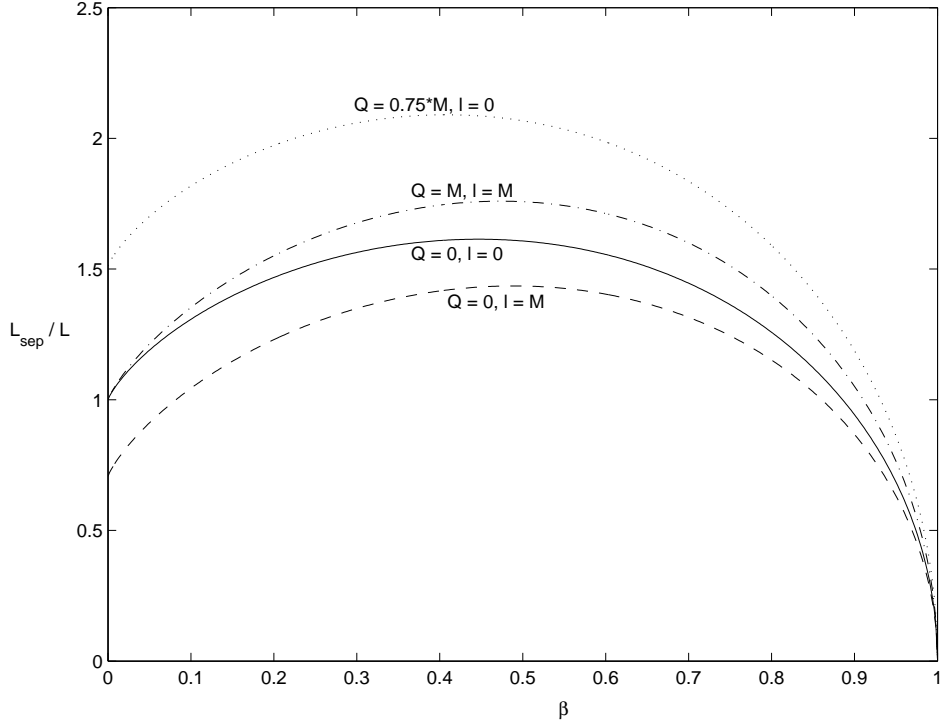}
  \caption{Proper distance between black hole poles as a function of
  the parameter $\beta$, for $\bar{u}\equiv 0$ and four particular
  choices of the quotients $Q/M$ and $l/M$.}  \label{fig:L_sep}
  \end{minipage}
\end{figure}

\begin{figure}[ht]
  \centering \begin{minipage}[t]{0.8\linewidth} \centering
  \includegraphics[width=0.95\textwidth, angle=0]{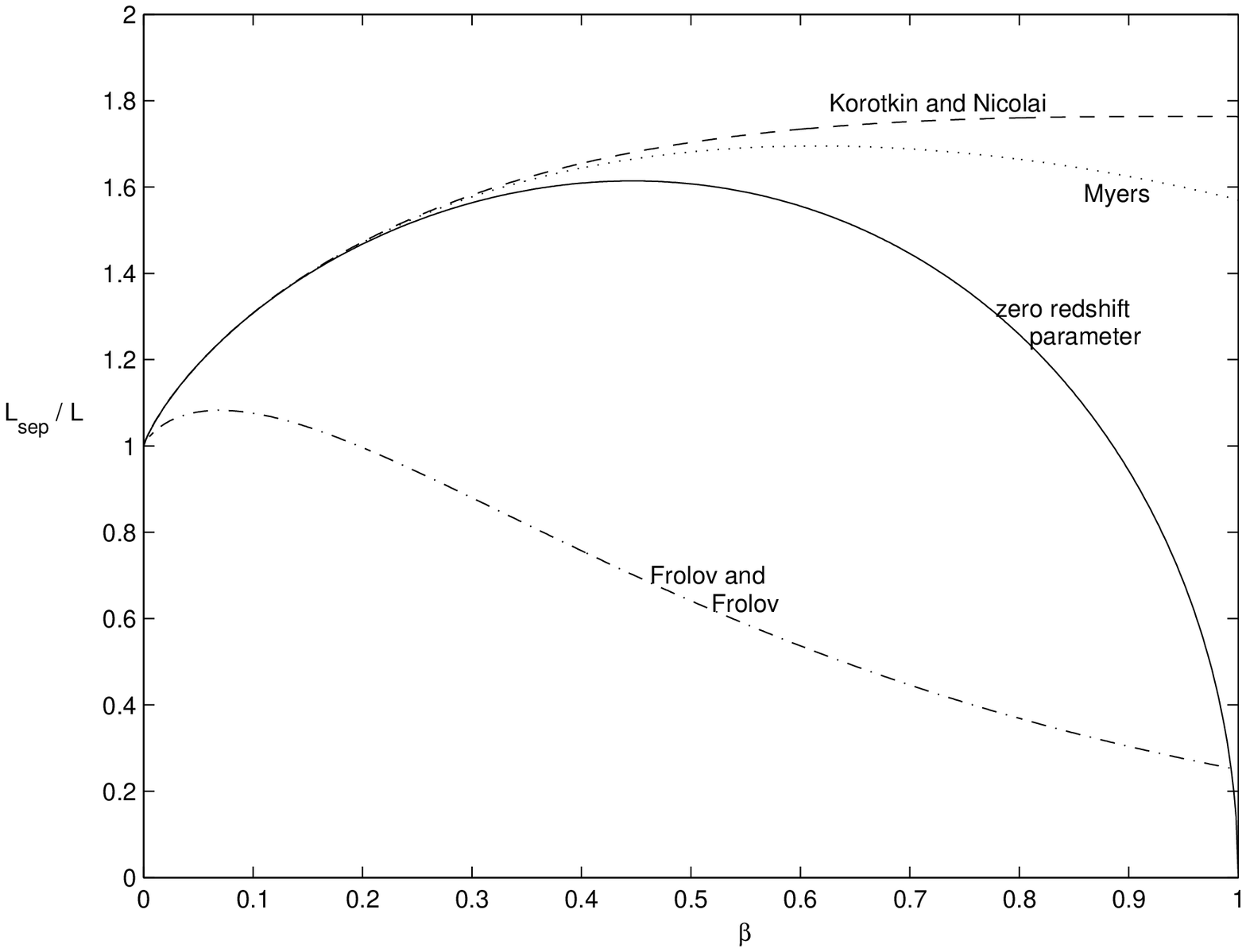}
  \caption{Proper distance between black hole poles as a function of
    the parameter $\beta$, for $Q = l = 0$ and four different choices
    of the redshift parameter $\bar{u}$, three of which being taken
    from the previous studies of the compactified Schwarzschild
    solution.}  \label{fig:L_sep2}
  \end{minipage}
\end{figure}

A quantity that characterizes the off-horizon section of the symmetry
axis is the proper spatial separation between the poles of the black
hole. It can be calculated as
\begin{equation}
  L_\mathrm{sep} = 2\Delta\int_1^{\beta^{-1}}\!\!\!f^{-1}dy,
\end{equation}
with $f$ evaluated at $x=0$. To get a general idea of how this
quantity depends on the choice of charges $M$, $Q$ and $l$, we have
held the quotients $Q/M$ and $l/M$ fixed at four different values and,
with the redshift parameter $\bar{u}$ set to zero, plotted
$L_\mathrm{sep}/L$ as a function of $\beta$ (figure \ref{fig:L_sep}).
Since $L_\mathrm{sep}$ depends nontrivially on how one chooses
$\bar{u}$ to depend on $\beta$, our plot for the case $Q = l = 0$
looks different than the one given by Frolov and Frolov \cite{Frolov}.
In particular, in \cite{Frolov} $L_\mathrm{sep}$ tends to a finite
value rather than zero as $\beta\rarr 1$, which is counter-intuitive
since in that limit the polar points reach each other as the event
horizon fills the whole axis. Again, this is directly related to the
behaviour of $\bar{u}$ in the same limit. However, any choice of
$\bar{u}$ which stays \emph{finite} results in a vanishing
$L_\mathrm{sep}$ as $\beta\rarr 1$, just like for $\bar{u} \equiv 0$.
To show the effect of different choices of $\bar{u}$, we have also
plotted $L_\mathrm{sep}$ as a function of $\beta$ for the different
choices of $\bar{u}$ collected in table \ref{table:ubar}, with the
choice $\bar{u} \equiv 0$ again included for comparison (figure
\ref{fig:L_sep2}).

\subsection{Large distance asymptotics}\label{sec:large}
In the limit $\rho\rarr\infty$, our compactified black hole solution
approaches an exact solution corresponding to the field of a
homogeneous line mass with electric as well as NUT line charge. This
solution is obtained simply by truncating all terms that fall off
exponentially with $\rho$, which is easy to do given the fourier
expansions derived in subsection \ref{sec:fourier}. Explicitly,
\begin{align}
  &\omegaAL = \bar{u}-\frac12\ln{[2\varphi_\beta(2)]}+\beta\ln{\!\parfrac{\rho}{2L}} \\
  &k = k_* + \beta^2\ln{\!\parfrac{\rho}{\Delta}} \\ \lbeq{Omegacyl}
  &\Omega = \frac{4lz}{L}+\Omega',
\end{align}
so the spacetime metric for our compactified Reisser-Nordstr\"om-NUT
solution, in the nonextremal case $\Delta>0$, rapidly approaches
\begin{equation}\lbeq{cylgen}
  ds^2 = -f^2(dt+\Omega d\phi)^2 + 
  f^{-2}\left[\tilde{C}^2(\rho/\rho_0)^{2\beta^2}(d\rho^2+dz^2)+\rho^2 d\phi^2\right],
\end{equation}
where
\begin{equation}\lbeq{F2cyl}
  f^2 = \frac{(r_+ - r_-)^2}{\left[r_-(\rho/\rho_0)^\beta-r_+(\rho/\rho_0)^{-\beta}\right]^2 + l^2\left[(\rho/\rho_0)^\beta-(\rho/\rho_0)^{-\beta}\right]^2} \\
\end{equation}
As before, $r_\pm = M\pm\Delta$ with $\Delta = \sqrt{M^2+l^2-Q^2}$,
but we have here also introduced the constants
\begin{align}
  &\rho_0 = \left[2\varphi_\beta(2)\,e^ {-2\bar{u}}\right]^{\frac1{2\beta}}2L \\
  &\tilde{C} = \left[\frac{\varphi_\beta(2)\,e^{2\bar{u}}}2\right]^{-\frac{\beta}2}e^{\beta^2+\int_0^\beta\ln[\varphi_{\beta'}(2)]d\beta'}.
\end{align}
If we set both the electric charge $Q$ and NUT charge $l$ to zero, the
function $f^2$ simplifies to
\begin{equation}
  f^2 = \parfrac{\rho}{\rho_0}^{\!\!2\beta},
\end{equation}
which shows that this is a special case of Levi-Civita's static
cylindrically symmetric vacuum solution. As discussed for instance by
Bi\u{c}\'ak \etal\cite{Bicak:2004fw}, the Levi-Civita metric contains two
essential parameters $m$ and $\mathcal{C}$ and can be written in the
standard form
\begin{equation}
  ds^2 = -\check\rho^{2m}d\check{t}^2+\check\rho^{2m(m-1)}(d\check\rho^2+d\check{z}^2)+\frac1{\mathcal{C}^2}\check\rho^{2(1-m)}d\phi^2, 
\end{equation}
where the coordinates $\check{t}$, $\check\rho$ and $\check{z}$ as
well as the \emph{conicity parameter} $\mathcal{C}$ in general carry
dimension in a nonstandard way. The conicity parameter can obviously
be set to unity by rescaling the coordinate $\phi$ as
$\phi\rarr\mathcal{C}\phi$, but since we insist that $\phi$ be a
periodic coordinate with the standard range $\phi\in[0,2\phi)$, the
value of $\mathcal{C}$ cannot be viewed as a coordinate choice. In
fact, since we are dealing with a compactified spacetime for which $z$
is also a periodic coordinate with a fixed range, the general static
cylindrically symmetric solution -- what one might call the
\emph{compactified Levi-Civita solution} -- has instead three
essential parameters, which for instance can be taken to be our
$\beta$, $\rho_0$ and $\tilde{C}$, with $\beta$ directly corresponding
to $m$.

Setting the electric charge $Q$ to zero while having a non-vanishing
NUT charge $l$ makes it possible to reexpress $f^2$ as
\begin{equation}
  f^2 = \frac{2\Delta}{r_+(\rho/\rho_0)^{2\beta}-r_-(\rho/\rho_0)^{-2\beta}} = \frac{\Delta}{l}\frac1{\cosh{[2\beta\ln{(\rho/\rho_1)}]}}, 
\end{equation}
where (note that $r_-<0$ in this case)
\begin{equation}
  \rho_1 = \left(-\frac{r_-}{r_+}\right)^{\!\!\frac1{4\beta}}\rho_0. 
\end{equation}
As should be expected, this vacuum solution is the ``cylindrical
analogue of NUT space'' that was constructed and studied by
Nouri-Zonoz\cite{Nouri-Zonoz:1997ms}. According to Nouri-Zonoz, the
``gravitomagnetic charge per unit length'' is (up to sign) equal to
the coefficient of $z$ in the expression for $\Omega$,
\ie in our case $4l/L$. However, what we have constructed here is the 
asymptotic field of a black hole with NUT charge $l$ in a space with a
compactified dimension of ($z$-coordinate) length $L$, so we expect
the NUT charge per unit $z$-length to be $l/L$. We believe that the
discrepancy of a factor of $4$ is only a matter of definitions (\ie
Nouri-Zonoz' ``gravitomagnetic charge'' is four times NUT charge, when
the latter is definied so that its value for the RNN metric
\refeq{rnn} is $l$.

Setting $l=0$ but keeping a non-zero electric charge $Q$, the general
asymptotic solution \refeq{cylgen} instead becomes the cylindrically
symmetric Einstein-Maxwell solution with a purely electric Maxwell
field. This is one of the solutions constructed and studied by
Richterek \etal \cite{Richterek:2000dh} as well as Miguelote \etal
\cite{Miguelote:2000qi}. The function $f^2$ can in this case be written as
\begin{equation}
  f^2 = \frac{\Delta^2}{Q^2}\frac1{\sinh^2{\![\beta\ln{(\rho/\rho_2)}]}}, 
 \qquad \rho_2 = \parfrac{r_+}{r_-}^{\!\!\frac1{2\beta}}\rho_0. 
\end{equation}
Note that $f^2$ diverges at a finite cylinder radius $\rho=\rho_2$,
indicating that there should be a curvature singularity there. This is
indeed the case, as noticed in
\cite{Richterek:2000dh,Miguelote:2000qi}, but let us see this
explicitly by calculating the energy density of the electromagnetic
field with respect to the unit timelike vector field $u^a =
f^{-1}\xi^a$ where $\xi^a = (\partial/\partial t)^a$, as before. The
result, in the general case with all three charges $M$, $Q$ and $l$
arbitrary, is
\begin{equation}
  \mu_\mathrm{EM} = u^a u^b T_{ab} = \frac12 u^a u^b G_{ab} = 
  2\parfrac{Q}{\tilde{C}\rho_0 L}^{\!\!2}(\rho/\rho_0)^{-2(1+\beta^2)}f^4.
\end{equation}
Clearly $\mu_\mathrm{EM}$ diverges at finite $\rho$ exactly when $f^2$
does. However, the curvature singularity at $\rho=\rho_2$ disappears
as soon as a NUT charge $l$ is turned on, since the denominator of the
right hand side of eq.\ \refeq{F2cyl} is then the sum of two squares
which cannot vanish simultaneously. Of course, one can make the
curvature grow arbitrarily large close to $\rho=\rho_2$ by choosing
the quotient $l/Q$ sufficiently small. This feature is not only
present in the asymptotic solution discussed here, but also in the
full compactified RNN solution with the $z$-dependent fourier terms
turned back on. The reason for this is that the range of the function
$\omega$ changes from $\omega\in(-\infty,0)$ to
$\omega\in(-\infty,\infty)$ when we compactify the spacetime by making
the coordinate $z$ periodic; In both cases $\omega\rarr-\infty$ occurs
at the event horizon, \ie at $\rho=0$, $\abs{z}\leq\Delta$, but
$\rho\rarr\infty$ sends $\omega$ to $+\infty$ rather than zero in the
compactified case. In particular, for $Q\neq 0 = l$, the compactified
$\omegaAL$ will take on the positive value
\begin{equation}
  \omegaAL_\mathrm{crit} = \frac12\ln{\!\!\parfrac{r_+}{r_-}},
\end{equation}
for which $f^2$ diverges, as can be seen from eq.\ \refeq{Fgeod}.
Again, this gives a curvature singularity which however is regularized
by turning on a NUT charge $l$, however small. The difference,
compared to the asymptotic solution, is that the curvature singularity
now occurs on some torus $\rho=f(z)$ which is not a simple flat torus
$\rho=\mathrm{constant}$. The asymptotic solution also has a curvature
singularity on the axis $\rho=0$ for $\beta\in(0,1)$, but this is in
contrast to the full compactified RNN solution which by construction
is regular on that axis. Another interesting feature related to the
NUT parameter $l$ is that although the spacetime metric \refeq{cylgen}
is cylindrically symmetric in the sense that no physically measurable
quantities depend on the $z$, this coordiate still appears in the
expression \refeq{Omegacyl} for $\Omega$. What is interesting is that
if we demand that the metric be explicitly periodic in $z$ with period
$L$, then the metric must be unchanged under $\Omega'\rarr\Omega'+4nl$
for any integer $n$. This is true, granted that $t$ is periodic with
period $8\pi l$ (just like for the non-compactified RNN solution we
started out with), since we can mimick the shift in $\Omega'$ by
letting $t\rarr t+4nl\phi$.

Turning finally to the extremal case $\Delta=0$, we must use $\omega =
M\omegaAL/\Delta$ instead of $\omegaAL$ before taking the limit
$\Delta\rarr 0$. The limit then gives
\begin{align}
  &\omega = 
\frac{2M}{L}\left[u' + \gamma + \ln{\!\parfrac{\rho}{2L}}\right], \qquad u' = \lim_{\beta\rarr 0}\beta^{-1}\bar{u} \\
  &k = 0 \\
  &\Omega = \frac{4lz}{L} + \Omega',
\end{align}
so the asymptotic spacetime metric is in this case
\begin{equation}
  ds^2 = -f^2(dt+\Omega d\phi)^2 + f^{-2}\left(d\rho^2+dz^2+\rho^2 d\phi^2\right),      
\end{equation}
with, using eq.\ \refeq{Fextr}, 
\begin{equation}\lbeq{Fextrasympt}
  f^2 =
\frac1{\left[1-\dsfrac{2M}{L}\,\ln{\!(\rho/\rho_3)}\right]^2+\left[\dsfrac{2l}{L}\,\ln{\!(\rho/\rho_3)}\right]^2},
 \qquad \rho_3 = 2Le^{-u'-\gamma}. 
\end{equation}
Just like in the nonextremal case, this asymptotic solution fails to
be asymptotically flat, albeit in a weaker manner. Also, the
discussion concerning the curvature singularity at finite $\rho$
applies here as well. In particular, from eq.\ \refeq{Fextr} it
follows that if $l=0$, the asymptotic solution has a curvature
singularity at $\rho=\rho_3 e^{L/(2M)}$, but the singularity is
removed when the NUT charge $l$ is turned back on.

\section{Conclusions}
Having constructed and analyzed what we consider to be the natural
periodic analogue of the Reissner-Nordstr\"om-NUT solution, a question
that naturally arises is whether or not any given asymptotically flat
black hole has a periodic analogue and, if so, whether or not the
periodic analogue is unique, given the period $L$ of the compactified
coordinate. By identifying the target space geodesic that the
non-compactified solution defines and assuming that the very same
geodesic should be used for the compactified version, we have found
that the problem essentially reduces to finding the appropriate
axisymmetric solution to the Laplace equation in flat space, just as
in the Schwarzschild case. There is thus a simple superposition
principle at play, which makes compactification tractable when Weyl
coordinates are used. However, the low dimension requires a
regularization procedure which introduces a one-parameter ambiguity as
there is no obvious way to fix the freedom to add an arbitrary
constant to the solution to the Laplace equation. In all previous
works on the periodic Schwarzschild solution, this constant -- the
redshift parameter -- was always fixed at an early stage and without
any physical motivation for the specific choice made. As we have seen
there are three different choices in the existing papers on the
subject, which we take as support for instead keeping the parameter
free, or at least fixing it at a later stage from physical
considerations. One way of fixing it would be to require that a set of
relations that hold for horizon quantities (for instance the relation
between the horizon area and the black hole charges) should remain the
same after compactification. If one could motivate why this should be
the case, our method would give a neat one-to-one correspondance
between standard and periodic Reissner-Nordstr\"om-NUT black holes. Of
course, we have not given any proof that our method is correct, in
some suitably defined sense. One could perhaps argue that it is not,
based on the fact that our solution family has a curvature singularity
surrounding the horizon in the case of electric but no NUT charge.
However, we see no way of avoiding such a singularity. Indeed, the
singularity in question is actually a general feature not just for our
full solution with a nontrivial dependence on the periodic coordinate,
but also for the cylindrically symmetric solution that is approached
for large cylinder radii. Since there simply is no other cylindrically
symmetric solution available that could work as an asymptotic solution
for a compactified Reissner-Nordstr\"om black hole, it seems at least
likely that the exponentially decaying corrections to the asymptotic
solution cannot smooth out the singularity.

One can think of several ways to extend our work while staying in four
dimensions. One way would be to include a dilaton field with arbitrary
coupling parameter. This would be straightforward to do, as our method
would still be applicable with the dilaton simply entering as an extra
target space coordinate. It would be interesting to see whether the
inclusion of a dilaton could prevent the above-mentioned singularity
from occuring. A general framework for distorted charged dilaton black
holes has been provided by Yazadjiev\cite{Yazadjiev:2000by}.

It would be even more interesting if the periodic analogue of the Kerr
solution could be constructed. Some indications of how this could be
done were given in \cite{Korotkin2}, but no explicit formulae were
given there, except for a few first steps. Since the noncompactifed
Kerr solution traces out a target space two-surface rather than a
geodesic, it would no longer be possible to linearize the problem and
thus more sophisticated methods, such as B\"acklund transformations,
are needed.

It would of course be even more interesting if our work could give
some insights into how things work in higher dimensions. Consider the
Einstein-Maxwell equations for a static $(d+1)$-dimensional spacetime.
Dimensionally reducing with respect to the static Killing vector, one
arrives at Einstein gravity in $d$ euclidean dimensions, coupled to a
two-dimensional (three-dimensional if a dilaton is included) sigma
model. One question one could ask is whether or not the target space
geodesic that the $(d+1)$-dimensional Reissner-Nordstr\"om solution
corresponds to would also be traced out when the solution is
compactified on a circle. The work of Harmark and Obers actually
suggests that this could be the case, as their ansatz is such that the
target space coordinates are functionally dependent, thus
corresponding to geodesic solutions. This does not mean that it would
be easy to make analytical progress for $d>3$, but nevertheless it
could give a coordinate invariant way of thinking about the ansatz of
these authors which in turn could lead to a deeper understanding of
the general problem. It would be very interesting to investigate this
matter further.

\section*{Acknowledgements}
The work of RvU was supported by the Ministry of Education of the Czech Republic under the project MSM 0021622409. The work of MK was supported by a post doctoral fellowship of the Faculty of Science of Masaryk University.


\begin{thebibliography}{99}

\bibitem{Majumdar}
S.~D.~Majumdar,
``A Class Of Exact Solutions Of Einstein's Field Equations,''
Phys.\ Rev.\  {\bf 72} (1947) 390.

\bibitem{Papapetrou}
A.~Papapetrou,
Proc.\ R.\ Irish\ Acad.\ {\bf A51}, (1947) 191.

\bibitem{Myers}
R.~C.~Myers,
``Higher Dimensional Black Holes In Compactified Space-Times,''
Phys.\ Rev.\ D {\bf 35} (1987) 455.

\bibitem{Korotkin1}
D.~Korotkin and H.~Nicolai,
``A Periodic analog of the Schwarzschild solution,''
arXiv:gr-qc/9403029.

\bibitem{Korotkin2}
D.~Korotkin and H.~Nicolai,
``The Ernst equation on a Riemann surface,''
Nucl.\ Phys.\ B {\bf 429} (1994) 229
[arXiv:gr-qc/9405032].

\bibitem{Frolov}
A.~V.~Frolov and V.~P.~Frolov,
``Black holes in a compactified spacetime,''
Phys.\ Rev.\ D {\bf 67} (2003) 124025
[arXiv:hep-th/0302085].

\bibitem{GL1}
R.~Gregory and R.~Laflamme,
``Black strings and p-branes are unstable,''
Phys.\ Rev.\ Lett.\  {\bf 70} (1993) 2837
[arXiv:hep-th/9301052].

\bibitem{GL2}
R.~Gregory and R.~Laflamme,
``The Instability of charged black strings and p-branes,''
Nucl.\ Phys.\ B {\bf 428} (1994) 399
[arXiv:hep-th/9404071].

\bibitem{Horowitz}
G.~T.~Horowitz and K.~Maeda,
``Fate of the black string instability,''
Phys.\ Rev.\ Lett.\  {\bf 87} (2001) 131301
[arXiv:hep-th/0105111].


\bibitem{HO1}
T.~Harmark and N.~A.~Obers,
``Black holes on cylinders,''
JHEP {\bf 0205} (2002) 032
[arXiv:hep-th/0204047].

\bibitem{HO2}
T.~Harmark and N.~A.~Obers,
``New phase diagram for black holes and strings on cylinders,''
Class.\ Quant.\ Grav.\  {\bf 21} (2004) 1709
[arXiv:hep-th/0309116].

\bibitem{HO3}
T.~Harmark and N.~A.~Obers,
``Phase structure of black holes and strings on cylinders,''
Nucl.\ Phys.\ B {\bf 684} (2004) 183
[arXiv:hep-th/0309230].

\bibitem{HO4}
T.~Harmark
``Small black holes on cylinders,''
Phys.\ Rev.\ D {\bf 69} (2004) 104015
[arXiv:hep-th/0310259].

\bibitem{HO5}
H.~Elvang, T.~Harmark and N.~A.~Obers,
``Sequences of bubbles and holes: New phases of Kaluza-Klein black holes,''
JHEP {\bf 0501} (2005) 003
[arXiv:hep-th/0407050].

\bibitem{Gubser}
S.~S.~Gubser,
``On non-uniform black branes,''
Class.\ Quant.\ Grav.\  {\bf 19} (2002) 4825
[arXiv:hep-th/0110193].

\bibitem{Kol1}
B.~Kol,
``Topology change in general relativity and the black-hole black-string
transition,''
arXiv:hep-th/0206220.

\bibitem{Kol2}
B.~Kol and T.~Wiseman,
``Evidence that highly non-uniform black strings have a conical waist,''
Class.\ Quant.\ Grav.\  {\bf 20} (2003) 3493
[arXiv:hep-th/0304070].

\bibitem{Kol3}
B.~Kol, E.~Sorkin and T.~Piran,
``Caged black holes: Black holes in compactified spacetimes. I: Theory,''
Phys.\ Rev.\ D {\bf 69} (2004) 064031
[arXiv:hep-th/0309190].

\bibitem{Kol4}
E.~Sorkin, B.~Kol and T.~Piran,
``Caged black holes: Black holes in compactified spacetimes. II:
5d numerical implementation,''
Phys.\ Rev.\ D {\bf 69} (2004) 064032
[arXiv:hep-th/0310096].

\bibitem{Kol5}
D.~Gorbonos and B.~Kol,
``A dialogue of multipoles: Matched asymptotic expansion for caged black
holes,''
JHEP {\bf 0406} (2004) 053
[arXiv:hep-th/0406002].

\bibitem{Kol6}
D.~Gorbonos and B.~Kol,
``Matched asymptotic expansion for caged black holes: Regularization of the
post-Newtonian order,''
arXiv:hep-th/0505009.

\bibitem{Wiseman1}
T.~Wiseman,
``Static axisymmetric vacuum solutions and non-uniform black strings,''
Class.\ Quant.\ Grav.\  {\bf 20} (2003) 1137
[arXiv:hep-th/0209051].

\bibitem{Wiseman2}
T.~Wiseman,
``From black strings to black holes,''
Class.\ Quant.\ Grav.\  {\bf 20} (2003) 1177
[arXiv:hep-th/0211028].

\bibitem{Wiseman3}
H.~Kudoh and T.~Wiseman,
``Properties of Kaluza-Klein black holes,''
Prog.\ Theor.\ Phys.\  {\bf 111} (2004) 475
[arXiv:hep-th/0310104].

\bibitem{Wiseman4}
H.~Kudoh and T.~Wiseman,
``Connecting black holes and black strings,''
Phys.\ Rev.\ Lett.\  {\bf 94} (2005) 161102
[arXiv:hep-th/0409111].

\bibitem{Emparan}
R.~Emparan and H.~S.~Reall,
``Generalized Weyl solutions,''
Phys.\ Rev.\ D {\bf 65} (2002) 084025
[arXiv:hep-th/0110258].

\bibitem{Fairhurst:2000xh}
S.~Fairhurst and B.~Krishnan,
``Distorted black holes with charge,''
Int.\ J.\ Mod.\ Phys.\ D {\bf 10} (2001) 691
[arXiv:gr-qc/0010088].

\bibitem{Geroch}
R.~Geroch and J.~B.~Hartle,
``Distorted black holes,''
Jour.\ Math.\ Phys.\  {\bf 23} (1981) 680

\bibitem{Bicak:2004fw}
J.~Bicak, T.~Ledvinka, B.~G.~Schmidt and M.~Zofka,
``Static fluid cylinders and their fields: global solutions,''
Class.\ Quant.\ Grav.\  {\bf 21} (2004) 1583
[arXiv:gr-qc/0403012].

\bibitem{Richterek:2000dh}
L.~Richterek, J.~Novotn\'y and J.~Horsk\'y,
``New Einstein-Maxwell fields of Levi-Civita's type,''
Czech.\ J.\ Phys.\  {\bf 50} (2000) 925
[arXiv:gr-qc/0003004].

\bibitem{Miguelote:2000qi}
A.~Y.~Miguelote, M.~F.~A.~da Silva, A.~Wang and N.~O.~Santos,
``Levi-Civita solutions coupled with electromagnetic fields,''
Class.\ Quant.\ Grav.\  {\bf 18} (2001) 4569
[arXiv:gr-qc/0104018].

\bibitem{Nouri-Zonoz:1997ms}
M.~Nouri-Zonoz,
``Cylindrical analogue of NUT space: Spacetime of a line gravomagnetic
monopole,''
Class.\ Quant.\ Grav.\  {\bf 14} (1997) 3123
[arXiv:gr-qc/9706015].


\bibitem{Yazadjiev:2000by}
S.~S.~Yazadjiev,
``Distorted charged dilaton black holes,''
Class.\ Quant.\ Grav.\  {\bf 18} (2001) 2105
[arXiv:gr-qc/0012009].

\end{thebibliography}
\end{document}